\documentclass{article}

\usepackage[final]{neurips_2019_ml4ps}

\usepackage[utf8]{inputenc} % allow utf-8 input
\usepackage[T1]{fontenc}    % use 8-bit T1 fonts
\usepackage{hyperref}       % hyperlinks
\usepackage{url}            % simple URL typesetting
\usepackage{booktabs}       % professional-quality tables
\usepackage{amsfonts}       % blackboard math symbols
\usepackage{amsmath}
\usepackage{amssymb}
\usepackage{graphicx}
\usepackage{subcaption}
\usepackage{nicefrac}       % compact symbols for 1/2, etc.
\usepackage{microtype}      % microtypography
\usepackage{color}
\usepackage{tikz}

\newcommand{\vc}[0]{\boldsymbol}  % Vector

\newcommand{\reals}[0]{\mathbb R}

\newcommand{\maybeequals}{\stackrel{?}{=}}

\DeclareMathOperator*{\argmax}{arg\,max}

\newcommand{\scriptD}[0]{\mathcal D}

\newcommand{\scriptL}[0]{\mathcal L}
\newcommand{\scriptM}[0]{\mathcal M}
\newcommand{\scriptO}[0]{\mathcal O}
\newcommand{\scriptU}[0]{\mathcal U}

\setcitestyle{numbers}

\title{Data-driven discovery of free-form governing differential equations}

\author{%
  Steven Atkinson\thanks{GE Research, Niskayuna, NY, USA.} \\
  \texttt{steven.atkinson1@ge.com} \\
  \And
  Waad Subber\footnotemark[1] \\
  \texttt{waad.subber@ge.com}\\
  \And
  Liping Wang\footnotemark[1] \\
  \texttt{wangli@ge.com}\\
  \AND
  Genghis Khan\footnotemark[1]\\
  \texttt{khan@ge.com}
  \And
  Philippe Hawi\thanks{University of Southern California, Los Angeles, CA.}\\
  \texttt{hawi@usc.edu}
  \And
  Roger Ghanem\footnotemark[2] \\
  \texttt{ghanem@usc.edu}
}

\begin{document}

\maketitle

\begin{abstract}
	We present a method of discovering governing differential equations from data without the need to specify \textit{a priori} the terms to appear in the equation.
	The input to our method is a dataset (or ensemble of datasets) corresponding to a particular solution (or ensemble of particular solutions) of a differential equation.
	The output is a human-readable differential equation with parameters calibrated to the individual particular solutions provided.
	The key to our method is to learn differentiable models of the data that subsequently serve as inputs to a genetic programming algorithm in which graphs specify computation over arbitrary compositions of functions, parameters, and (potentially differential) operators on functions. 
	Differential operators are composed and evaluated using recursive application of automatic differentiation, allowing our algorithm to explore arbitrary compositions of operators without the need for human intervention.
	We also demonstrate an active learning process to identify and remedy deficiencies in the proposed governing equations.
\end{abstract}

\section{Introduction}

The modern scientist enjoys access to ever-growing amount of laboratory data to help uncover new knowledge.
However, this abundance of data can be unwieldy to analyze using traditional methods of deduction from which governing laws usually arise.
Motivated by the notion of machine learning as a partner in the scientific process,
we introduce a method to automate the process of understanding and manipulating data for the purpose of hypothesizing, criticizing, and ultimately discovering novel physical governing laws.
Our work is similar in spirit to a growing literature on machine learning and differential programming to solve differential equations 
\cite{tsoulos2006solving, izzo2017differentiable, long2017pde, de2018solving}, 
and calibration of physical laws where the involved terms are known \textit{a priori}
\cite{bongard2007automated, schmidt2009distilling, ly2012learning, brunton2016discovering, raissi2017machine, raissi2017physics1, raissi2017physics2, rudy2017data, schaeffer2017learning, raissi2018hidden, raissi2018deep, raissi2018multistep, rudy2019data, lee2019coarse}.

Here, we focus on discovering \textit{free-form} equations in that, unlike previous work, we do not require an \textit{a priori} specification of the terms that may appear in the equation to be discovered; instead we take as input a library of primitive (algebraic and differential) operators on functions from which expressions are composed, calibrated, and evaluated on the fly.
The result is a method that allows substantial flexibility in discovering differential equations while still producing a highly structured result (a compositional expression represented as a graph) conducive to further study and analysis.

\section{Methodology}
\label{sec:methodology}
\begin{figure}[hbt]
	\centering
	\includegraphics[width=0.8\textwidth]{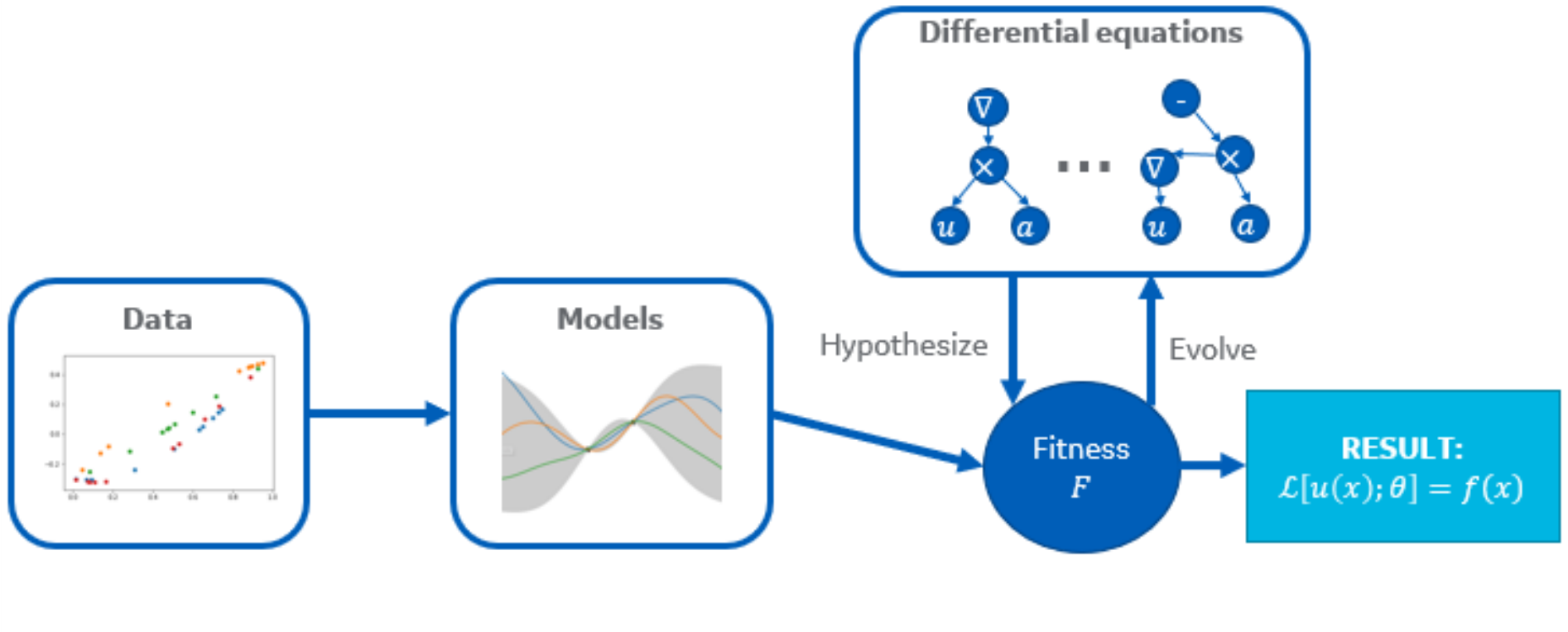}
	\caption{
		Schematic of our methodology for discovering governing differential equations from data.
	}
	\label{fig:methodology:overview}
\end{figure}

Consider the space of pairs of functions 
$\scriptU_\scriptL = \{ (\vc u: \Omega_x \rightarrow \Omega_u \subseteq \reals^{d_u}, f: \Omega_x \rightarrow \reals) \}$ 
such that 
$\scriptL[\vc u] = f ~ \forall (\vc u, f) \in \scriptU_\scriptL$, 
where $\Omega_x \subset \reals^{d_{st}}$ is a $d_{st}$-dimensional spatiotemporal domain of interest.
Some (possibly stochastic) observation operator $\scriptO: \scriptU \rightarrow \scriptD$ produces discrete samples of these functions as a data set $\vc D_i = \{\vc d_{ij} \}_{j=1}^{d_u+1}$ (e.g.\ sensor digital readouts).
Given $N$ ``experiments'' $\{ \vc D_i \}_{i=1}^N$ from possibly-distinct $(\vc u, f)$ pairs, we seek to determine the differential equation $\scriptL$ obeyed within our class of physical systems of interest $\scriptU_\scriptL$ as well as to calibrate any parameters in $\scriptL$ particular to the data observed.
We present a methodology, summarized schematically in Fig.\ \ref{fig:methodology:overview}, to solve this problem, explained presently.

The first step to our approach is to fit explicit, differentiable models $\{ M_{ij} \}_{i, j=1}^{N, d_u}$, $M_{ij} \in \scriptM$ to each scalar output data set $\{ \vc d_{ij} \}$. 
We considered fully-connected feedforward neural networks \cite{raissi2018deep} and Gaussian processes\footnote{We use the posterior mean of the GP model in subsequent manipulations.}.
Note that the model architectures (e.g.\ choice of nonlinearities and kernels) must be selected to reflect the desired differentiability properties matching our inductive biases about the nature of $\scriptU_\scriptL$.

\begin{figure}[h]
	\centering
	\begin{subfigure}{0.3\linewidth}
		\centering
\begin{tikzpicture}
\node at (0, 2) {$\nabla_{\vc x} \cdot ()$};

\draw [->] (0, 1.7) -- (0, 1.3);
\node at (0, 1) {$\times$};

\draw [<->] (-0.7, 0.3) -- (0, 0.7) -- (0.7, 0.3);
\node at (-1, 0) {$-()$};
\node at (1, 0) {$\nabla_{\vc x}$};

\draw [->] (-1, -0.3) -- (-1, -0.7);
\node at (-1, -1) {$a(\vc x)$};

\draw [->] (1, -0.3) -- (1, -0.7);
\node at (1, -1) {$u(\vc x)$};
\end{tikzpicture}
		\caption{
			The elliptic operator: \\
			$\nabla \cdot \left(-a(\vc x) \nabla u(\vc x)\right)$.}
		\label{figure:graph-examples:elliptic}
	\end{subfigure}
	\begin{subfigure}{0.3\linewidth}
		\centering
\begin{tikzpicture}
\node at (-1, 3) {$+$};

\draw [<->] (-1.7, 2.3) -- (-1, 2.7) -- (-0.3, 2.3);
\node at (-2, 2) {$\partial_t$};
\node at (0, 2) {$+$};

\draw [->] (-2, 1.7) -- (-2, 1.3);
\node at (-2, 1) {$\partial_t$};

\draw [->] (-2, 0.7) -- (-2, 0.3);
\node at (-2, 0) {$x(t)$};

\draw [<->] (-0.7, 1.3) -- (0, 1.7) -- (0.7, 1.3);
\node at (-1, 1) {$\partial_t$};
\node at (1, 1) {$x(t)$};

\draw [->] (-1, 0.7) -- (-1, 0.3);
\node at (-1, 0) {$x(t)$};
\end{tikzpicture}
		\caption{
			Second-order ordinary differential operator: \\
			$\ddot{x}(t) + \dot{x}(t) + x(t)$.
		}
		\label{figure:graph-examples:ode}
	\end{subfigure}
	\begin{subfigure}{0.3\linewidth}
		\centering
\begin{tikzpicture}
\node at (1, 4) {$-$};
\draw [<->] (0.3, 3.3) -- (1, 3.7) -- (1.7, 3.3);

\node at (0, 3) {$\partial_t$};
\draw [->] (0, 2.7) -- (0, 2.3);
\node at (0, 2) {$\partial_t$};
\draw [->] (0, 1.7) -- (0, 1.3);
\node at (0, 1) {$u(\vc x, t)$};

\node at (2, 3) {$\times$};
\draw [<->] (1.3, 2.3) -- (2, 2.7) -- (2.7, 2.3);

\node at (1, 2) {$\theta_1$};

\node at (3, 2) {$\nabla_{\vc x} \cdot()$};
\draw [->] (3, 1.7) -- (3, 1.3);
\node at (3, 1) {$\nabla_{\vc x}$};
\draw [->] (3, 0.7) -- (3, 0.3);
\node at (3, 0) {$u(\vc x, t)$};

\end{tikzpicture}
		\caption{
			The wave equation operator: \\
			$\ddot u (\vc x, t) - \theta_1 \nabla^2 u(\vc x, t)$
		}
		\label{figure:graph-examples:wave}
	\end{subfigure}
	\caption{
		Examples of compositional differential operators represented as graphs combining functions with primitive algebraic and differential operators.
	}
	\label{figure:graph-examples}
\end{figure}
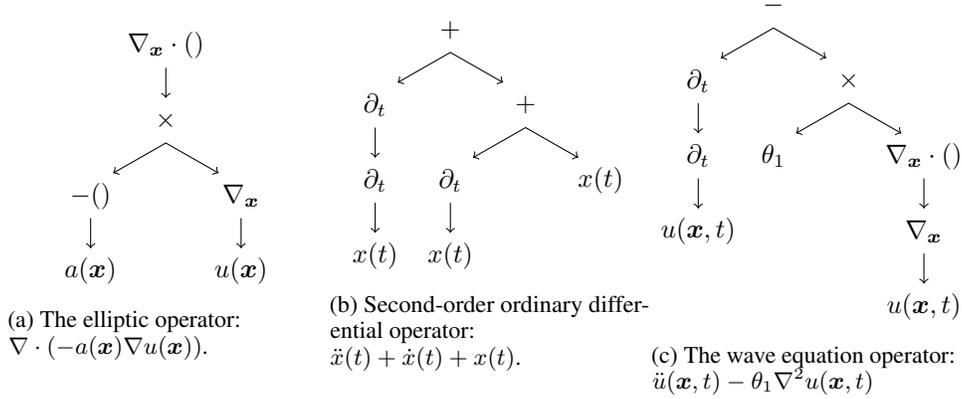

Second, having equipped our data $\{\vc d_{ij} \}$ with differentiable function representations $\{ M_{ij}\}$, we utilize a genetic programming algorithm to compose differential equations.
Let a differential equation (hereafter referred to as an individual) be represented by a tree graph $G(V,E)$ whose leaves are instances of the fitted models and parent verticies are $n$-ary operators selected from a user-supplied library of operators.
We do not restrict ourselves to algebraic operators $(+,-,\times,\dots)$, but include \textit{differential} operators as well $(\partial/\partial x_i, \nabla(\cdot), \nabla \cdot (\cdot),\dots)$. 
Each leaf of the graph template is primarily associated with models of a particular output dimension $j$, but \textit{realizations} of the graph instantiate all of the leaves from some ``experiment'' $i$.
Differential operators are computed using automatic differentiation using PyTorch \cite{paszke2017automatic};
Critically and in contrast to prior work, arguments to the operator nodes are functions, not arrays of numbers, allowing us to compose $G$ as a \textit{function} to evaluate at any point in $\Omega_x$ to return a residual $r(\vc x)$.
Figure \ref{figure:graph-examples} provides some example expressions represented as graphs that such as those that our method handles.

Finally, parameters $\vc \theta$ to be calibrated enter $G$ as leaves representing constant functions; we calibrate them in a Bayesian manner using black-box variational inference \cite{ranganath2014black}.
Reflecting our ignorance about the role of the $\{\theta_i\}$'s in $G$, we use a flat prior for $p(\vc \theta)$ and non-factorized multivariate Gaussian variational posterior $q(\vc \theta)$.
The likelihood is a Gaussian $r(\vc x)$ with point estimate variance.

The hypothesis space over graphs is explored via genetic programming \cite{banzhaf1998genetic}; 
we use the evolutionary algorithm implemented by deap \cite{fortin2012deap} to initialize, mutate, and mate graphs. 
Individuals compete using the evidence lower bound (ELBO) evaluated at the inputs associated with the $\{\vc D_i \}$'s as the fitness function $L$.
By iterating the EA, differential equations are automatically hypothesized and criticized; the outcome of the EA is a list of candidate differential equations $(G_1, G_2, \dots)$ ranked by their fitness so $L(G_1, D) \ge L(G_2, D) \ge \dots$, reflecting the degree to which they (through the fit models $\vc M$) explain the observed data.
Minimality may be encouraged through multi-objective optimization with a second fitness function favoring parsimony as in \cite{clune2013evolutionary, gaier2019weight} but was not considered in our current work as we did not require it to discover the correct underlying equations in our examples.

If one has the ability to gather more data, then our approach may be paired with an active learning loop to sequentially obtain additional samples to augment the $D_i$'s using $a(\vc x) = r^2(\vc x)$ under $G_1$ as an acquisition function.
This causes one to look more closely where the best explanation of the observed data is most inadequate.
One may also define a tolerance $\delta$ such that when $a^* = \argmax_{\vc x \in \Omega_x} a(\vc x) < \delta$, the iterative algorithm is said to have converged and the discovered equation explains the observed data to a user-specified degree.

\section{Examples}
\label{sec:examples}
Code for examples in this section will be made public upon publication.
In all of the following examples, we use the posterior means of Gaussian process models with squared exponential kernels \cite{rasmussen2006gaussian} to fit functions to the data;
Fully-connected neural networks tended not to capture the gradients of the data accurately.
We hypothesize this can be attributed to inferior inductive biases conferred by typical architectures, and is a subject of ongoing investigation.
The evolutionary algorithm uses a population of $512$ individuals, probability of mutation $0.2$, and probability of mating $0.8$.
These hyperparameters were not fine-tuned as we did not find it necessary in order to obtain satisfactory performance.

\subsection{Synthetic benchmarks}
\label{sec:benchmarks}

We benchmark the above methodology by applying it to a number of classic differential equations enumerated in Table \ref{table:examples}.
We select ground truths with unity coefficients to focus here on discovering the structure of the equation; calibration of parameters will be emphasized in the example of Sec.\ \ref{sec:ultrasound}.
Data for the ODEs and one-dimensional elliptic problems were produced using Python's SciPy library and FEniCS \cite{logg2012automated}, respectively.
For the two-dimensional elliptic equation, data was obtained from \cite{atkinson2019structured}.
We repeat each experiment $40$ times with different random number generator seeds to understand the robustness of our method.

\begin{table}[ht]
	\begin{center}
		\caption{
			Summary of the benchmark differential equations discovered using our methodology.
		}
		\label{table:examples}
		\begin{tabular}{rrlr}
			Differential equation & Differential operator & Dependent variables & $f \maybeequals 0$
			\\ \hline
			ODE & $\ddot{u} + \dot{u} + u$ & $u(t): [0, 10] \rightarrow \reals$ & Yes \\
			Heterogeneous elliptic PDE & $\frac{d}{dx} (-a(x) \frac{du}{dx})$ & $a: [0, 1] \rightarrow \reals^+$, $u(x): [0, 1] \rightarrow \reals$ & No \\
			Nonlinear elliptic PDE & $\frac{d}{dx} (-a(u) \frac{du}{dx})$ & $u(x): [0, 1] \rightarrow \reals$ & No \\
			2D Heterogeneous elliptic PDE & $\nabla \cdot (-a(x) \nabla u)$ & $a: [0, 1]^2 \rightarrow \reals^+$, $u(x): [0, 1]^2 \rightarrow \reals$ & Yes \\
		\end{tabular}
	\end{center}
\end{table}

\begin{figure}[hbt]
	\centering
	\begin{subfigure}{0.45\linewidth}
		\centering
		\includegraphics[width=\linewidth]{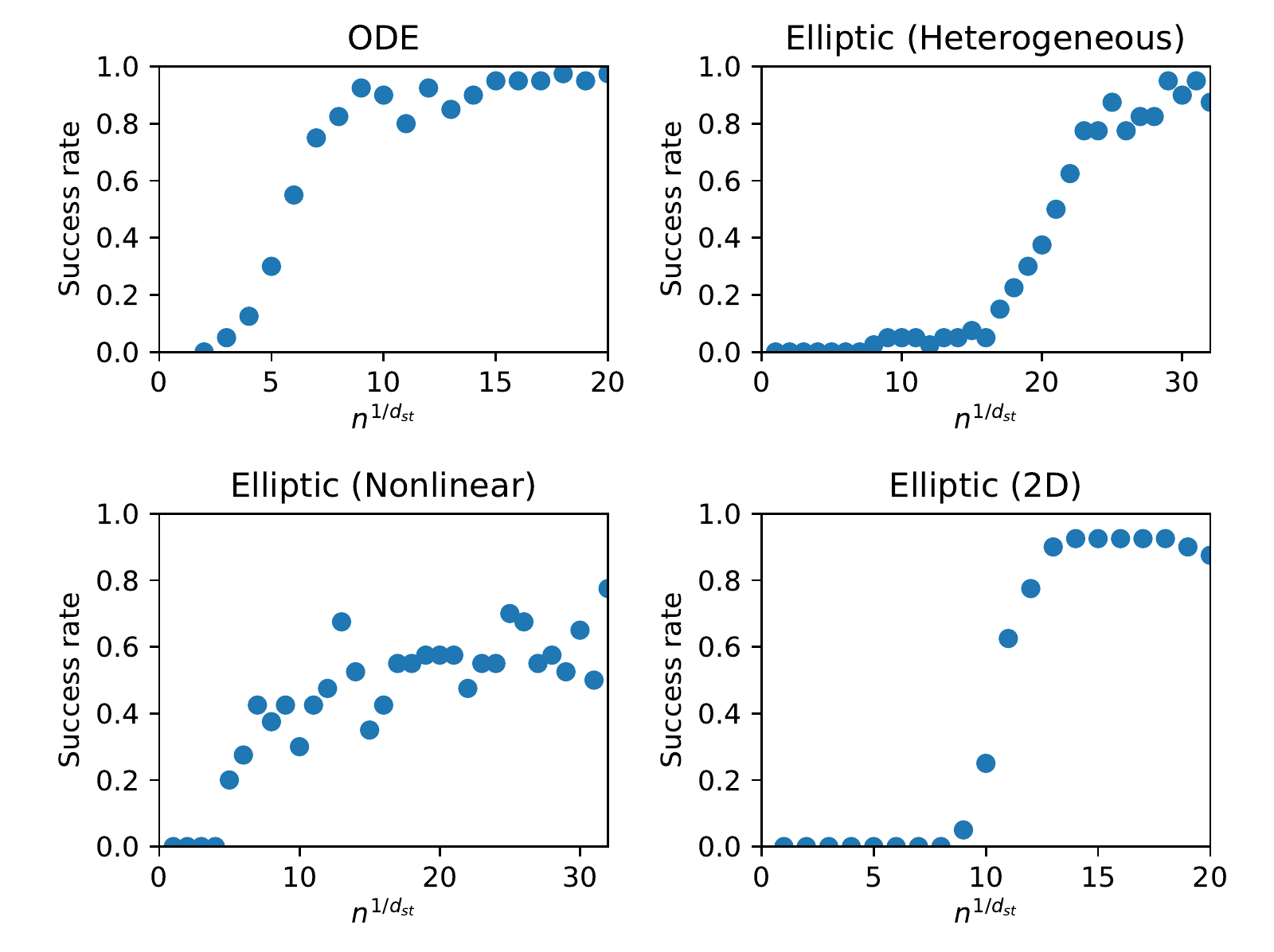}
		\caption{Success rates}
		\label{figure:success-rate}
	\end{subfigure}
	\begin{subfigure}{0.45\linewidth}
		\centering
		\includegraphics[width=\linewidth]{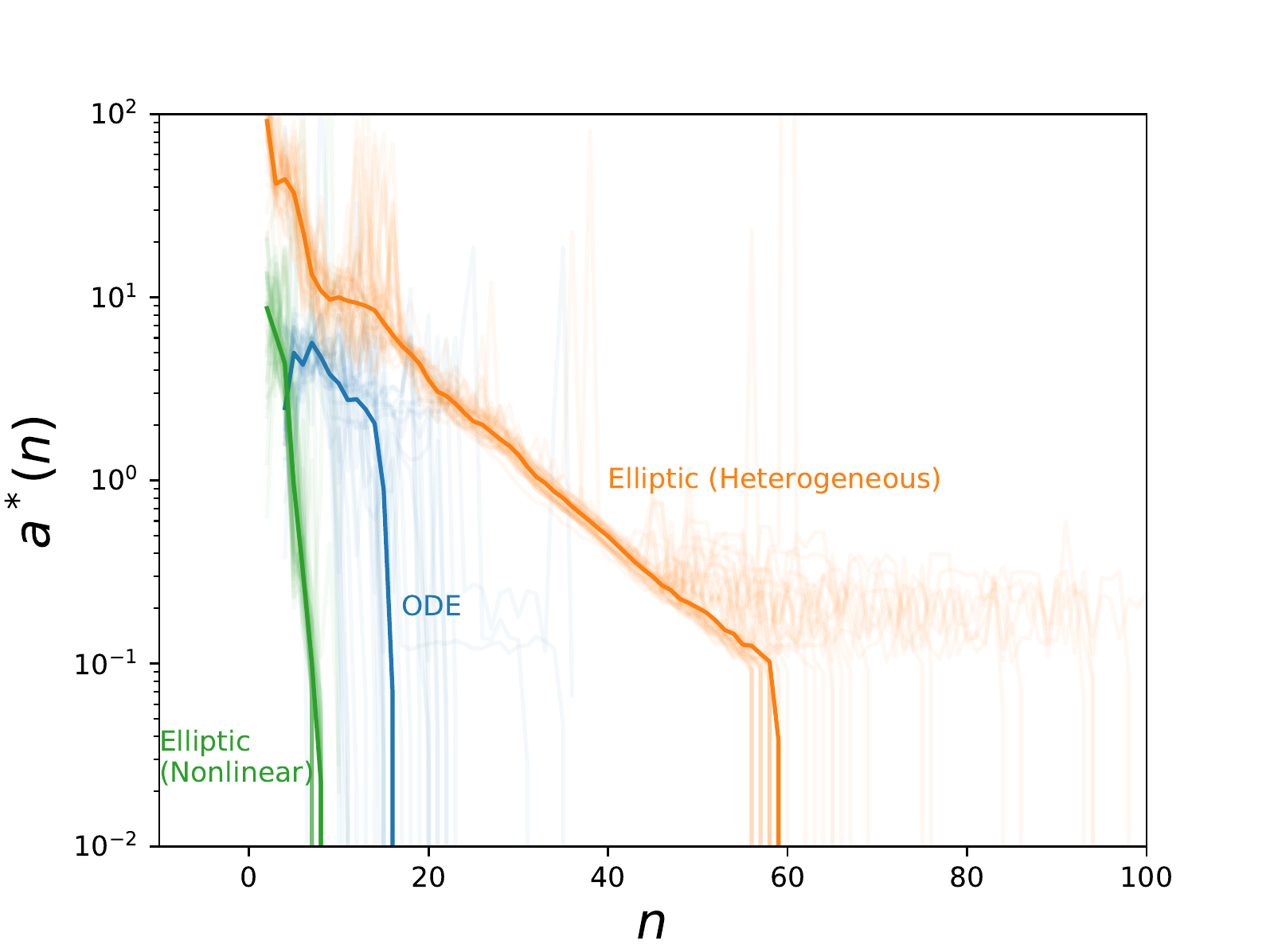}
		\caption{Adaptive sampling}
		\label{figure:adaptive}
	\end{subfigure}
	\caption{
		(\subref{figure:success-rate}) Frequency of success in discovering the underlying differential equations for the benchmark problems as a function of $n^{1/d_{st}}$.
		(\subref{figure:adaptive}) Maximum discrepancy as quantified by $a^*$ during adaptive acquisition of data.
		Different curves show repeats for a given problem.
		Heavy lines show the median performance over 40 repeats.
	}
\end{figure}

Figure \ref{figure:success-rate} shows the success rate for the benchmark systems listed in Table \ref{table:examples} as the number of samples available from the experiment $n$ is varied.
We report results in terms of $n^{1/d_{st}}$, where $d_{st}$ is the number of independent (spatiotemporal) variables in the problem.
Figure \ref{figure:adaptive} shows the performance of our method in the adaptive setting in which data are added sequentially following Sec.\ \ref{sec:methodology}, as quantified by the evolution of $a^*$ with increasing $n$.
We see that a termination tolerance of $\delta=0.1$ seems sufficient to ensure the discovery of the correct equations, and many runs consistently identify the correct equation far in advance of termination.

\subsection{Real-world ultrasound experiment}
\label{sec:ultrasound}
\begin{figure}[hbt]
	\centering
	\includegraphics[width=\linewidth]{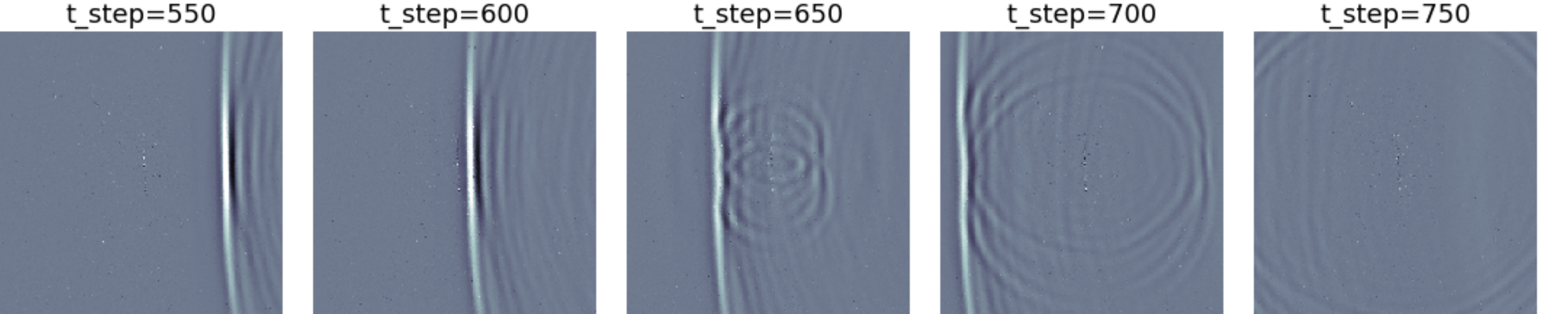}
	\caption{
		Snapshots of the ultrasound experiment at various time steps.
	}
	\label{figure:ultrasound}
\end{figure}

Finally, we demonstrate our approach on discovering the physics of a laboratory ultrasound experiment.
A laser is used to measure the deflection of a cracked Aluminum alloy specimen subjected to an acoustic impulse.
The wave travels through the material and exhibits back-scattering at the crack location (Fig.\ \ref{figure:ultrasound}).

It is expected based on expert knowledge that the data should be well-described by the wave equation, $\scriptL^w[u] = u_{tt} - \alpha \Delta^2 u_{xx} = 0$, throughout the bulk of the homogeneous material.
However, this fails to capture inhomogeneities, dissipative forces, and out-of-plane behavior that might be amended with corrective terms.
It is of interest to build a computational model based on the discovered physics from which we may ultimately solve the inverse problem of inferring unseen crack morphologies from a sparse sensor array.
We restrict ourselves to searching for a corrector $\scriptL'$ such that 
$u_{tt} - \scriptL[u; \vc \theta] = 0$, 
with 
$\scriptL = \theta_1 \nabla^2 u +  \scriptL'[u; \vc \theta_{2:}]$. 
Parameters $\vc \theta$ are calibrated as explained in Sec.\ \ref{sec:methodology}.

\begin{figure}[hbt]
	\centering
	\includegraphics[width=\linewidth]{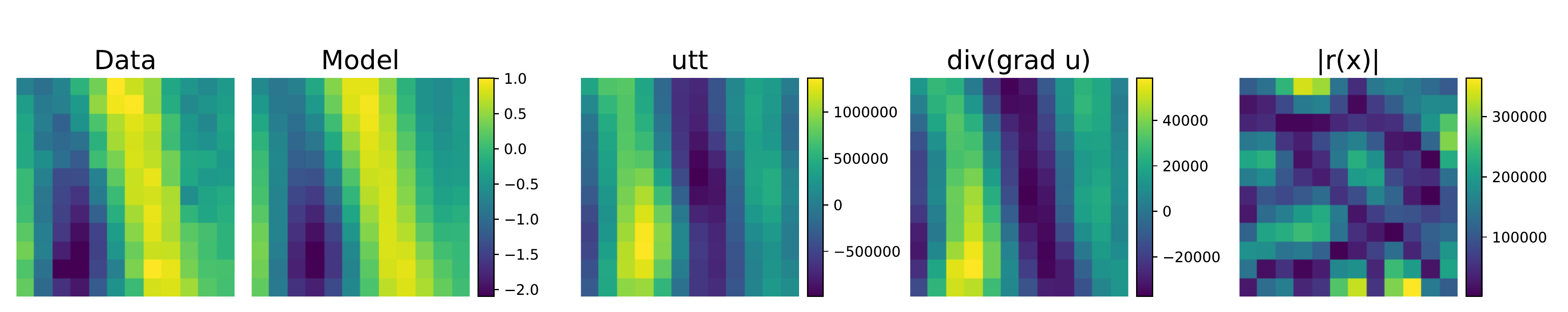}
	\caption{
		From left to right: 
		A slice in time of the ultrasound data,
		its GP function representation used in exploring differential equations,
		the term $u_{tt}(x)$ evaluated on the function,
		the term $\nabla^2 u$ evaluated on the function,
		and the residual of the calibrated wave equation using the mode of $q(\vc \theta)$.
	}
	\label{figure:ultrasound-modeled}
\end{figure}

We preprocess the data by cropping to a smaller window in spacetime to exclude data where no significant activity is observed; the resulting data are fitted with a GP.
Table \ref{table:ultrasound} enumerates the fittest correctors discovered from the first 32 proposed individuals along with the mean and standard deviation of the parameters' marginal posteriors and fitness (ELBO) associated with the individuals.
Figure \ref{figure:ultrasound-modeled} illustrates a slice of the dataset along with its GP model, the differential terms $u_{tt}$ and $\nabla^2 u$, and the resulting residual associated with the calibrated wave equation using the mode of $q(\vc \theta)$.

\begin{table}[ht]
	\begin{center}
		\caption{
			Potential correctors discovered for the ultrasound dataset.
			The uncorrected equation is provided last as a baseline.
		}
		\label{table:ultrasound}
		\begin{tabular}{rrr}
			$\scriptL[u]$ & $q(\vc \theta)$ (mean, std) & ELBO
			\\ \hline
			$\theta_1 \nabla^2 u + \theta_2 + \theta_3 u$ & 
			$(21.7, 0.14)$, $(120, 2300)$, $(-69, 2900)$ & 
			$\mathbf{-6.17445 \times 10^4}$
			\\
			$\theta_1 \nabla^2 u + \theta_2 u + \theta_3 u^2$ & 
			$(21.7, 0.14)$, $(-69, 3400)$, $(-3.78, 4.70)$ & 
			$-6.17450 \times 10^4$
			\\
			$\theta_1 \nabla^2 u + \theta_2 + \theta_3 u^2$ & 
			$(21.7, 0.14)$, $(110, 2300)$, $(83, 2400)$ & 
			$-6.17452 \times 10^4$ 
			\\
			$\theta_1 \nabla^2 u + \theta_2 u^2 + \theta_3 u^3$ & 
			$(21.7, 0.14)$, $(84, 2200)$, $(-45, 1400)$ & 
			$-6.17455 \times 10^4$
			\\
			$\theta_1 \nabla^2 u + \theta_2 u + \theta_3 u_t$ & 
			$(21.7, 0.14)$, $(-69, 3400)$, $(-3.78, 4.70)$ & 
			$-6.17509 \times 10^4$ 
			\\	
			$\theta_1 \nabla^2 u$ & 
			$(21.7, 0.14)$ & 
			$-6.17626 \times 10^4$
			\\
		\end{tabular}
	\end{center}
\end{table}

\vspace{-0.3in}

\section{Conclusions}  % and Discussion}
\label{sec:conclusion}
We have described a novel method for discovering governing differential equations with arbitrary structure from raw data and demonstrated its effectiveness on a number of standard differential equations.
Our method paves the way for an artificial intelligence ``research assistant'' as a companion to scientists seeking to understand novel physics.
Furthermore, the output of our method, being human-readable differential equations, is compatible with existing workflows available to engineers and scientists such as theoretical analysis and numerical simulation, including so-called physics-informed machine learning methods.

\subsubsection*{Acknowledgments}
The authors thank the United States Air Force Research Laboratory for providing the ultrasound data analyzed in Sec.\ \ref{sec:ultrasound} for public release under Distribution A as defined by the United States Department of Defense Instruction 5230.24.
This material is based upon work supported by the Defense Advanced Research Projects Agency (DARPA) under Agreement No.\ HR00111990032.
Approved for public release; distribution is unlimited.

\bibliographystyle{unsrt}
% Added from bbl:

\end{document}